\documentclass[aps,prl,reprint,amssymb,showpacs,superscriptaddress,notitlepage,twocolumns]{revtex4-1}
\usepackage{bm}
\usepackage{graphicx}
\usepackage{dcolumn}
\usepackage{braket}
\usepackage{natbib}
\usepackage{amsmath}
\usepackage{color}
\usepackage{ulem}
\usepackage{siunitx}

\definecolor{gray}{rgb}{0.7,0.7,0.7}
\definecolor{orange}{rgb}{1, 0.4, 0}
\definecolor{dgreen}{rgb}{0.0, 0.4, 0.0}
\definecolor{yblue}{rgb}{0.06, 0.3, 0.57}

\newcommand\ldsout{\bgroup\markoverwith{\textcolor{blue}{\rule[0.5ex]{2pt}{0.4pt}}}\ULon}

\begin{document}
\title{Dipolar-induced enhancement of parametric effects in polariton waveguides}
\author{D. G. Su\'arez-Forero}
\affiliation{CNR NANOTEC, Institute of Nanotechnology, Via Monteroni, 73100 Lecce, Italy}
\affiliation{Dipartimento di Fisica, Universit\`{a} del Salento, Strada Provinciale
Lecce-Monteroni, Campus Ecotekne, Lecce 73100, Italy}
\author{F. Riminucci}
\affiliation{Dipartimento di Fisica, Universit\`{a} del Salento, Strada Provinciale
Lecce-Monteroni, Campus Ecotekne, Lecce 73100, Italy}
\affiliation{Molecular Foundry, Lawrence Berkeley National Laboratory, One Cyclotron Road, Berkeley, California, 94720, USA}
\author{V. Ardizzone}
\email{v.ardizzone85@gmail.com}
\affiliation{CNR NANOTEC, Institute of Nanotechnology, Via Monteroni, 73100 Lecce, Italy}
\author{N. Karpowicz}
\email{nicholas.karpowicz@nanotec.cnr.it}
\affiliation{CNR NANOTEC, Institute of Nanotechnology, Via Monteroni, 73100 Lecce, Italy}
\author{E. Maggiolini}
\affiliation{CNR NANOTEC, Institute of Nanotechnology, Via Monteroni, 73100 Lecce, Italy}
\author{G. Macorini}
\affiliation{CNR NANOTEC, Institute of Nanotechnology, Via Monteroni, 73100 Lecce, Italy}
\author{G. Lerario}
\affiliation{CNR NANOTEC, Institute of Nanotechnology, Via Monteroni, 73100 Lecce, Italy}
\author{F. Todisco}
\affiliation{CNR NANOTEC, Institute of Nanotechnology, Via Monteroni, 73100 Lecce, Italy}
\author{M. De Giorgi}
\affiliation{CNR NANOTEC, Institute of Nanotechnology, Via Monteroni, 73100 Lecce, Italy}
\author{L. Dominici}
\affiliation{CNR NANOTEC, Institute of Nanotechnology, Via Monteroni, 73100 Lecce, Italy}
\author{D. Ballarini}
\affiliation{CNR NANOTEC, Institute of Nanotechnology, Via Monteroni, 73100 Lecce, Italy}
\author{G. Gigli}
\affiliation{CNR NANOTEC, Institute of Nanotechnology, Via Monteroni, 73100 Lecce, Italy}
\author{A. S. Lanotte}
\affiliation{CNR NANOTEC, Institute of Nanotechnology, Via Monteroni, 73100 Lecce, Italy}
\affiliation{INFN, Sez. Lecce, Via per Monteroni, Lecce, Italy}
\author{K. West}
\affiliation{PRISM, Princeton Institute for the Science and Technology of Materials, Princeton Unviversity, Princeton, NJ 08540}
\author{L. Pfeiffer}
\affiliation{PRISM, Princeton Institute for the Science and Technology of Materials, Princeton Unviversity, Princeton, NJ 08540}
\author{D. Sanvitto}
\affiliation{CNR NANOTEC, Institute of Nanotechnology, Via Monteroni, 73100 Lecce, Italy}
\affiliation{INFN, Sez. Lecce, Via per Monteroni, Lecce, Italy}

\begin{abstract}
    Exciton-polaritons are hybrid light-matter excitations arising from the non-perturbative coupling of a photonic mode and an excitonic resonance. Behaving as interacting photons, they show optical third-order nonlinearities providing effects such as optical parametric oscillation or amplification. It has been suggested that polariton-polariton interactions can be greatly enhanced by inducing aligned electric dipoles in their excitonic part. However direct evidence of a true particle-particle interaction, such as superfluidity or parametric scattering is still missing. In this work, we demonstrate that dipolar interactions can be used to enhance parametric effects such as self-phase modulation in waveguide polaritons. By quantifying these optical nonlinearities we provide a reliable experimental measurement of the direct dipolar enhancement of polariton-polariton interactions. 
\end{abstract}
\maketitle

Adding high nonlinearities to compact, integrated optical circuits could unlock the development of new, fully photonic devices such as ultra-low-power active switches and transistors \cite{Ballarini2013}, parametric amplificators or oscillators and supercontinuum sources \cite{Walker2019}. Nonlinearities could even push forward the limit of integrated optics in quantum applications, i.e. universal quantum logic gates \cite{Slussarenko2019,Flamini2019}. Exciton-polaritons have demonstrated to be  good  candidates to supply nonlinearities to integrated optical platforms. They emerge, in inorganic semiconductors, from  the  strong  coupling  between  a  photonic  mode  and  an  electronic  excitation  (exciton) in Quantum  Wells  (QWs).  The nonlinearities in such systems, originating from short-range exchange interactions in the excitonic component, have been widely studied in the mesoscopic regime, allowing the demonstration of effects such as optical parametric oscillation, \cite{Baumberg2000}, bistability \cite{Baas2004}, superfluidity \cite{Amo2009,Lerario2017}, pattern formation \cite{Ardizzone2013} or quantum vorticity \cite{Lagoudakis2008}. Despite polariton-polariton interactions providing higher nonlinearities than standard nonlinear materials \cite{Walker2019}, they failed to bring such system fully in the quantum regime \cite{Sanvitto2016}. For example, in the recently observed polariton blockade, a consequence of the polaritonic interactions at the two-particles level \cite{Munoz-Matutano2019,Delteil2019}, only a weak  violation of the  classical regime has been observed, due to the low interaction over dissipation ratio. These results have  strongly  motivated the quest for the enhancement of the nonlinearities in polariton based systems. Recently, it has been given evidence that increasing the lengths of aligned exciton dipoles brings an enhancement of the polaritonic interactions in both vertical microcavities \cite{Togan2018}, and waveguides (WGs) \cite{Rosenberg2016, Rosenberg2018}, due to dipolar long range interactions between excitons. This very promising phenomenon could be exploited for many nonlinear  optical  effects or for the design of logic gates working at the two-particles regime \cite{Heuck2020}. However, up to now, an evidence that dipolar nonlinearities can actually work to enhance nonlinear parametric effects is still missing, with only indirect proofs that have been provided, as is the overall blueshift of the exciton resonance. Nevertheless, the overall blueshift is often due to a concurrence of different factors–such as phase space filling, charge screening, reduction of the Rabi coupling and presence of dark excitonic states–which can lead to puzzling results.
The WG geometry, a system in which the photonic mode is confined into a 2-D slab by total internal reflection \cite{Walker2013, Rosenberg2016}, seems particularly promising for the study and development of optoelectronic devices \cite{Suarez-Forero2020}, thanks to features such as faster propagating particles, a geometry that favors the integration of polaritonic optical circuits, high quality factors of the photonic mode and less demanding growth requirements. Most importantly, the WG platform allows the application of an external electric field perpendicular to the propagation direction able to control the polariton-polariton interaction strength through the manipulation of the exciton dipole length \cite{Rosenberg2018, Liran2018}. \\
In this letter, we demonstrate that dipolar polariton-polariton interaction do enhance parametric effects well beyond those derived by standard particle-particle exchange interaction. This is done by measuring the self phase modulation (SPM) of a resonant pulsed wavepacket propagating in the same kind of sample studied in \cite{Rosenberg2016, Rosenberg2018, Liran2018}. Moreover, we use this effect to make a precise and unequivocal assessment of the strength of the polariton dipolar interaction.\\

A short pulse propagating inside a polariton WG undergoes a parametric effect \cite{Walker2019} known as self-phase modulation (SPM). SPM is a well-known third order nonlinear effect  \cite{Agrawal2003} seen in optical systems as caused by the modification of the refractive index due to the propagation of the pulse itself and resulting in a spectral broadening and, for higher powers, structuring of the propagating pulse. However, the same effect can be seen, for strongly interacting particles, as a spontaneous parametric scattering, a four wave mixing process, that leads to depletion of the packet at the highest density point, in favour of a spread of particles along the propagation dispersion. 
Here we measure the SPM-induced pulse spectral modification in an AlGaAs/GaAs WG-QW system, directly showing the enhancement of the parametric effect by dipolar polariton-polariton interactions, introduced by means of a perpendicular electric field. The measured data are accurately fitted by a Gross-Pitaevskii (GP) model allowing exctracting the real dipolar enhancement of polariton-polariton interactions. Remarkably, we measure onto the parametric effect a 5-fold enhancement of excitonic interactions due to the change in the dipole length.

\begin{figure}
    \centering
    \includegraphics[width=1\columnwidth]{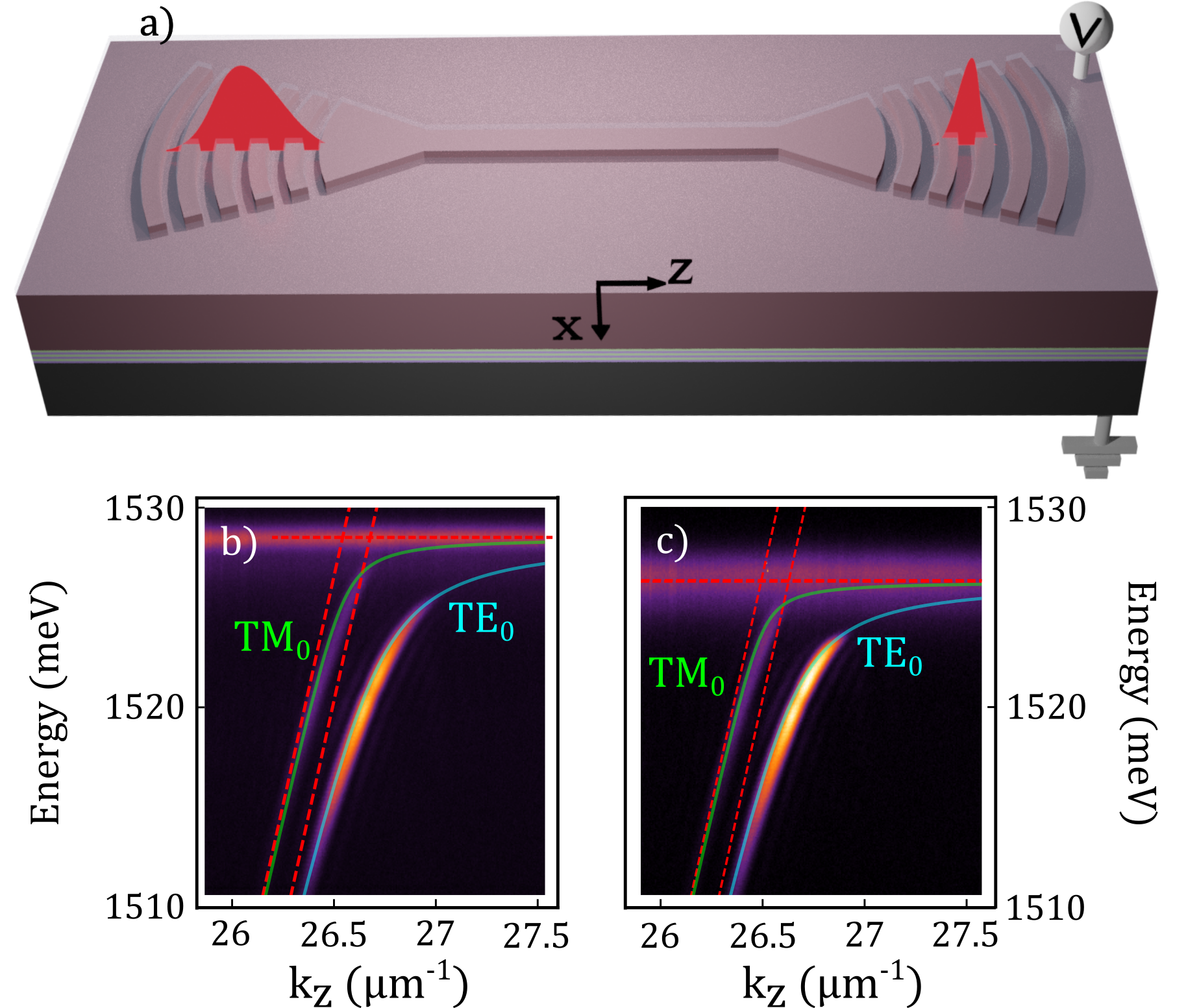}
    \caption{a) A slab waveguide is patterned through an etching process to create a wire $1$ $\mu m$ wide with input and output focusing gratings. The incident laser pulse resonantly injects population into the system, and the output signal, modified by the nonlinearities resulting from polaritonic interactions, is collected and analyzed. The doped substrate together with the ITO capping layer enable the application of an electric field along the growth direction.
    b) Polariton dispersion in the WG system, without applied electric field. c) As in b) for an applied electric field E = 11.9 $kV/cm$, resulting in the Stark shift of the excitonic energy.}
    \label{sketch}
\end{figure}

The WG structure is grown on an $n^+$-doped GaAs substrate, on top of which a $500$ nm cladding of $Al_{0.8}Ga_{0.2}As$ was previously deposited. The structure consists of 12 pairs of $20$ nm thick GaAs QW separated by $20$ nm of $Al_{0.4}Ga_{0.6}As$ barriers \cite{Rosenberg2016, Rosenberg2018}. A dry etching technique \cite{Liao2017} allows to process the slab WG in order to fabricate a $1$ $\mu m$ wide wire, confining polaritons in one additional direction and further reducing the modal volume. With this etching technique we also fabricate focusing gratings \cite{Mehta2017} or tapers to efficiently inject and extract the laser pulse into the small-volume waveguide. A schematic model of the WG system, comprising the input/output tapered gratings, is displayed in Fig.~\ref{sketch} a. Finally, the slab surface is covered with a $50$ $nm$ layer of Indium Tin Oxide (ITO), that together with the doped substrate on the opposite side, enables the appliance of an external electric field in the direction perpendicular to the slab plane \cite{Rosenberg2018} (see SI for additional details on sample etching and processing). Figure \ref{sketch} b shows the energy dispersion of the zero-order polaritonic modes in absence of external fields. As reported in previous works \cite{Rosenberg2016, Shapochkin2018}, the geometry and polarization of the guided electromagnetic modes yield a Rabi splitting around 3 times higher for the Transverse Electric (TE) than for the Tranverse Magnetic (TM) mode. The value of the Rabi splitting is $\Omega=13.9$ $meV$ for the TE mode and $\Omega=5.8$ $meV$ for the TM mode. Due to the greater exciton-TE mode coupling, all the measurements are performed on this mode. When an electric field is applied to the structure \cite{Rosenberg2016} the spatial profile of the potential energy of the QW is modified. The induced spatial separation between electrons and holes results in a finite dipole in a photocreated electron-hole pair \cite{Bastard1988,Bastard1983}. 
 As a consequence, the exciton energy shifts and the Rabi splitting slightly diminishes, as it is showed in figure \ref{sketch} c (see also SI for further details). The range of fields explored in this work corresponds to the regime characterized by a quadratic Stark energy shift \cite{Bastard1988}, in which exciton dissociation due to the applied field can be neglected. By using a variational approach \cite{Bastard1983}, we extract an induced dipole length around $4$ $nm$ for Fig.~\ref{sketch} c in agreement with previous works \cite{Rosenberg2018,Togan2018,Rosenberg2016}. The laser pulse is focused on the input taper, that injects the light into the TE$_0$ mode of the micro-wire. After propagating along $200$ $\mu m$ it is extracted by a second taper, see sketch in Fig.~\ref{sketch} a. The light is then collected and analyzed by a spectrometer, see SI for further details. During the pulse propagation, the time-dependent pulse intensity modifies the refractive index of the medium, due to ${\chi}^{(3)}$ nonlinearities provided by polariton-polariton interactions, resulting in the SPM effect. Figure \ref{sketch} a shows that the time-dependent modification of the refractive index can be intuitively understood as an induced temporal compression of the propagating pulse and hence the observed SPM spectral broadening. This effect can also be viewed as a particle-particle scattering mechanism: two particles injected with the same energy and wavevector can interact and scatter towards two different states fulfilling the energy and momentum conservation conditions \cite{Agrawal2003,Walker2019}. \\

\begin{figure}
    \centering
    \includegraphics[width=1\columnwidth]{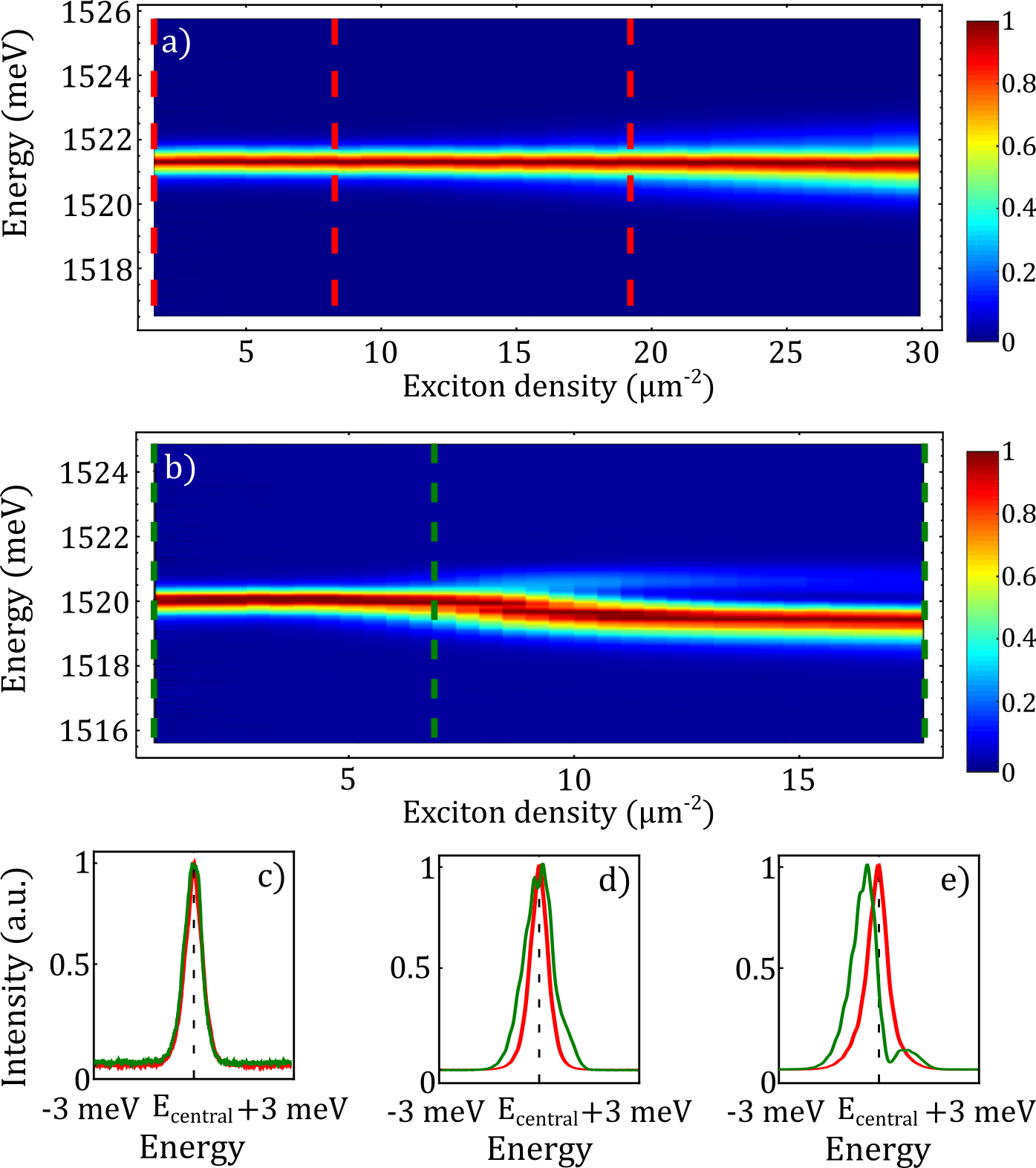}
    \caption{a) Normalized output pulse spectra as a function of the exciton densities. b) Same as panel a) but for an applied transverse electric field of $11.9$ $kV/cm$; nonlinear effects in b) are visible at exciton densities lower than in a), a signature of the dipolar enhancement of polariton - polariton interactions. c)-e) Pulse broadening and structuring for three different exciton densities indicated by vertical lines on the 2D maps of panels a) and b); the red spectra correspond to the case in which no electric field acts on the system, while green spectra correspond to an electric field value of $11.9$ $kV/cm$; spectra are energy shifted to compare shape and FWHM.}
    \label{spm}
\end{figure}
Figures \ref{spm} a-b show 2D representations of the pulse spectra obtained after resonant nonlinear propagation through the WG. The x axis represents the calculated exciton density (see description of the theoretical model below), which is proportional to the excitation power. Figure \ref{spm} a shows the propagation without applied electric field and Fig.~\ref{spm} b the propagation for $E = 11.9$ $kV/cm$. In both cases the laser energy is tuned to be at excitonic fraction $X^{2}=0.5$. Both figures show a density-dependent spectral broadening and structuring of the propagating pulse \cite{Agrawal2003,Walker2019}. 
\begin{figure}
    \centering
    \includegraphics[width=0.99\columnwidth]{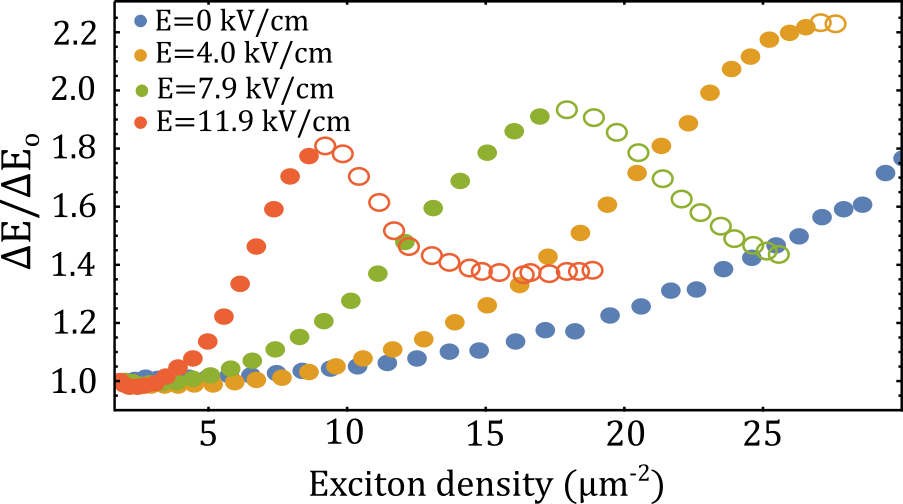}
    \caption{FWHM of the spectra from a gaussian fit as a function of the output power for four different values of the electric field. The output pulse width is normalized to the incident width. The empty circles correspond to the cases in which the output spectrum has several peaks, and represent the linewidths of the most intense lobe.}
    \label{fwhm}
\end{figure}
Remarkably, the pulse broadening in presence of an applied field in Fig.~\ref{spm} b is larger and occurs at lower exciton densities than in the case without applied electric field in Fig.~\ref{spm} a. Moreover, in presence of the applied electric field, the output pulse shows a more pronounced structure. At higher excitation powers, the high-intensity part (the central peak) of the propagating pulse is depleted by the particle-particle scattering resulting in a more complex structure of the output pulse \cite{Agrawal2003}. This is a signature of the nonlinear pulse propagation and can be understood in terms of a peak depletion due to the particle-particle scattering. Figure \ref{spm} also shows an asymmetric broadening, with the red part of the propagating pulse being more intense than the blue part. This is due to both the shape of the polariton dispersion and the increased absorption on the blue side due to the excitonic tail. 
We note that for E = 11.9 $kV/cm$, the output intensities are roughly half the output intensities without any applied electric field. We assign this discrepancy to a slight increase of nonradiative losses in presence of the electric field (see also discussion below). Figures \ref{spm} c-e show spectra of the output pulse at zero field (red lines) and at E = 11.9 $kV/cm$ (green lines) at exciton densities equal to 1, 7 and 19 excitons per $\mu m^{2}$, see corresponding vertical dashed lines in figures \ref{spm} a and b. These figures show that at low exciton density, see Fig.~\ref{spm} c, the nonlinearity does not affect the pulse shape and the two spectra have identical linewidths. When the exciton density is increased, Fig.~\ref{spm} d, the applied electric field causes a larger broadening of the output pulse. At even higher exciton densities, Fig.~\ref{spm} e, the pulse propagating under the action of the applied electric field shows a pronounced structure with the presence of two lobes, while the pulse propagating without applied electric field keeps a single peak shape.   
The effect of the electric field on the SPM-induced broadening is also clearly visible in Fig.~\ref{fwhm}, showing the FWHM of the output pulse as a function of the exciton densities for different applied electric fields (keeping the same excitonic fraction $X^{2}=0.5$ for each field). Filled points refer to the range of exciton densities in which the output pulse shows a single broadened peak. Empty circles refer to the range of exciton densities in which the output pulse show a structured peak (in this case, each point represents the FWHM of the more intense lobe). As the applied electric field increases, the pulse broadening at a given polariton density increases, showing a clear dipolar enhancement of the parametric effects. Moreover, pulse breaking, i.e. the transition from filled points to empty circles, takes place at lower excitonic density when the applied electric field is increased, highlighting the dipolar enhancement of polariton-polariton scattering.\\ 
To obtain a quantitative description of the observed dipolar enhancement of the polariton-polariton interactions, we solve the coupled equations \cite{Walker2019} \ref{GPE}a and \ref{GPE}b describing the system dynamics in the GP formalism: 

\begin{subequations}
\begin{eqnarray}
\left[i\frac{\partial}{\partial t}+i \gamma_p+\nu_g \left(i\frac{\partial}{\partial z}+\frac{1}{2k_z}\frac{\partial^2}{\partial x^2}\right) \right]A=\left(\frac{\Omega}{2}\right)\Psi,\\
\left[i\frac{\partial}{\partial t}+i \gamma_e-g|\Psi|^2 \right]\Psi=\left(\frac{\Omega}{2}\right)A,
\end{eqnarray}
\label{GPE}
\end{subequations}

\noindent where $A$ and $\Psi$ are the photon and exciton wavefunction amplitudes, respectively. $\nu_g$ is the particles group velocity obtained from the theoretically fitted dispersion and $k_z$ the corresponding linear momentum. $\gamma_p$ and $\gamma_e$ are the decay rates of the coherent photonic and excitonic states. The evolution of the initial ($\sim2-3$ $ps$)  pulse is calculated by solving the equations \ref{GPE} obtaining a theoretical output spectrum for each input power. 

\begin{figure}
    \centering
    \includegraphics[width=1\columnwidth]{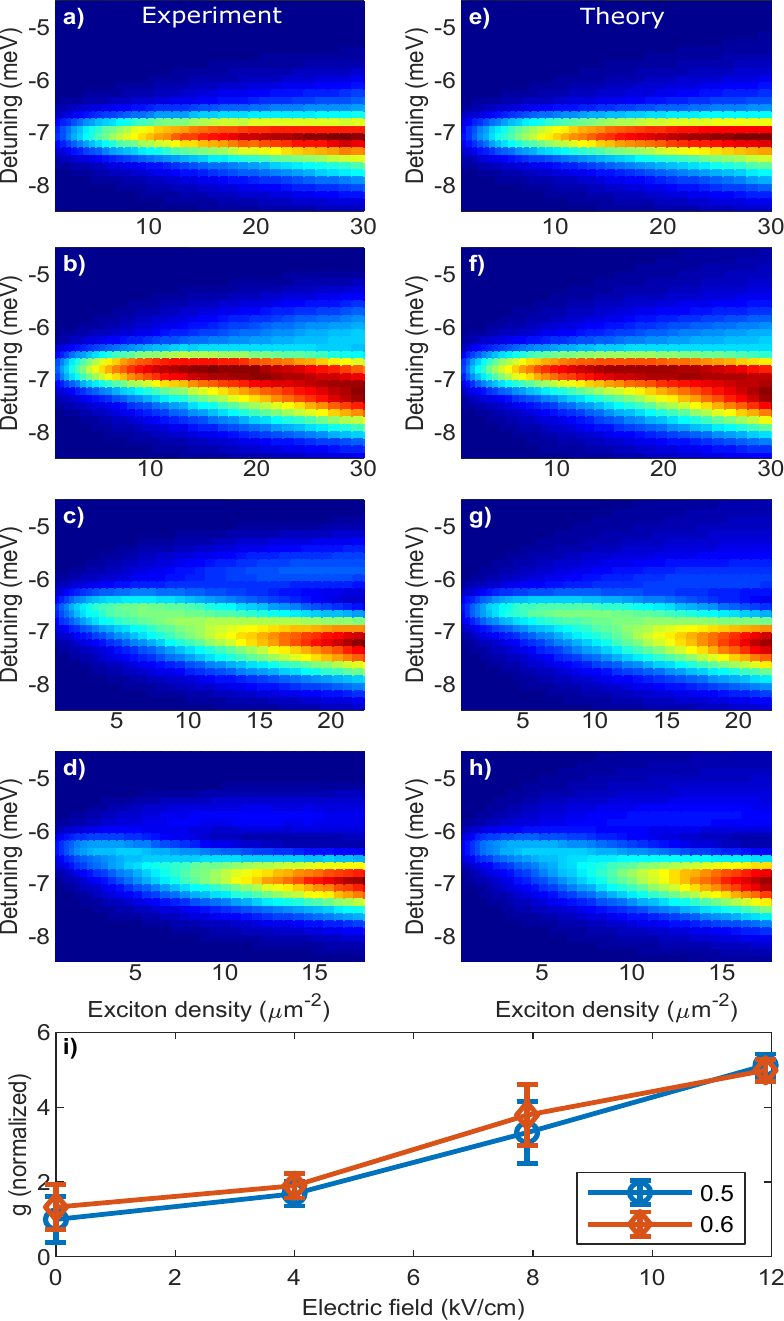}
    \caption{a-d) Measured density maps of the output spectra as a function of estimated exciton density at the end of the waveguide for field strengths of 0 $kV/cm$, 4 $kV/cm$, 7.9 $kV/cm$, and 11.9 $kV/cm$, respectively, and placed on a constantly spaced photon energy grid for comparison to simulations, for an excitonic fraction of 0.5. e-h) Corresponding results of numerical solution of the Gross-Pitaevskii equation for guided exciton-polaritons after nonlinear least squares fitting to determine the value of the interaction constant $g$. See the text for the calibration of the $|\Psi|^2$ axis. i) Electric field dependence of the fitted value of the parameter $g$, for two different excitonic fractions, 0.5 and 0.6, normalized to the value at zero field and fraction of 0.5.}
    \label{fit}
\end{figure}

A nonlinear least squares routine is used to determine $g$, as well as the detuning and duration of the pump pulse, minimizing the difference between the measured and simulated matrices of power vs. spectrum. Fig.~\ref{fit} shows the measured (left column) and calculated (right column) spectra as function of the input power for each of the four applied electric fields, with an excellent agreement between experimental data and simulations. It is worth noting that here we are focusing on the dipolar enhancement of polariton-polariton interactions. In particular, we have fixed the value of $g=6 ~ \mu eV ~ \mu m^{2}$ at zero applied electric field \cite{Estrecho2019} and found the corresponding exciton density per QW. Based on the obtained exciton density and on the measured output intensities, it is possible to retrieve the exciton density/output intensity calibration. We calibrate based on the peak value of $\left|\Psi\right|^2$ observed at the end of the waveguide in the simulation. We have then used this calibration to determine the exciton densities for the entire data set (see SI and in particular figure S3 for the details).   
Figure \ref{fit} i) shows that the enhancement of the dipolar interaction with the field causes $g$ to increase by a factor $\approx 5$ under the strongest electric field applied in the experiment. To show the consistency of our approach we have reported values of $g$ obtained on two different sets of data measured at excitonic fractions $X^{2} = 0.5$ and $X^{2} = 0.6$.\\

Remarkably, while we find a significant enhancement of polariton-polariton interactions due to the applied electric field, this enhancement is at least an order of magnitude smaller that the one reported on a sample having the same design \cite{Rosenberg2018} by measuring non-resonant blueshift of the polariton resonance. We believe, as mentioned above, that this discrepancy is essentially due to the fact that the polariton blueshift is only an indirect measurement of the particle-particle interaction that could be caused by many different effects. As a matter of fact, we have also performed non-resonant experiments under ultra-fast pumping, observing a wide range of possible blueshifts at a given excitation power. The cause of such behaviour is due to multiple factors not all related to the dipole enhancement and possibly dependent on extrinsic parameters. Non-resonant excitation could be responsible for the creation of trapped electric carriers that, under the action of the applied electric field and depending on the quality of the electric contacts, can induce a screening phenomenon. Such a screening may reduce the Stark effect and result in an apparent blueshift. In contrast, in this work we measure polariton-polariton interactions through a different effect, the SPM, which cannot be enhanced by other effects including screening. We stress also that the use of an ultrafast ps exciting pulse is important to reduce other spurious causes such as the creation of a long living, ``dark" exciton reservoir \cite{Estrecho2019}, that contribute to the overall density without being detected. Particularly in the WG platform, the presence of long-lived excitonic reservoir can alter the measurement of the interaction strength by more than two orders of magnitude \cite{Walker2015,Walker2017}. To avoid these possible spurious effects, we resonantly inject the population with a pulsed laser of $80$ MHz repetition rate and $\sim2-3$ $ps$ pulse duration. Our results are in qualitative agreement with the enhancement reported under resonant reflectivity in the case of permanent electric dipoles stemming from indirect excitons in double QWs \cite{Togan2018}.\\
  
In conclusion, we demonstrated that parametric effects in polaritonic systems can be enhanced by polarizing their excitonic component thanks to an applied electric field. By measuring the self-phase modulation of a propagating waveguide packet, we find a 5-fold enhancement of polariton-polariton interactions. However, in case of systems with very weak nonlinearities, such a dipolar effect, which is independent of the intrinsic polariton exchange interaction, could even lead to enhancements order of magnitude higher. For instance, we believe that this enhancement method could be extremely effective in TMD-based polariton systems. These results are essential to assess the relevance of dipolar polariton waveguides for future integrated and quantum optics application \cite{Suarez-Forero2019, Cuevas2018} and to guide the development of a new generation of WG polaritons supporting stronger nonlinearities, even at room temperature.\\

\noindent \textbf{Acknowledgments}\\
We thank Paolo Cazzato for technical support.\\
We are grateful to Ronen Rapaport for inspiring discussions and for sharing information about the sample design.\\
The authors acknowledge the project PRIN Interacting Photons in Polariton Circuits – INPhoPOL (Ministry of University and Scientific Research (MIUR), 2017P9FJBS\_001)\\
Work at the Molecular Foundry was supported by the Office of Science, Office of Basic Energy Sciences, of the U.S. Department of Energy under Contract No. DE-AC02-05CH11231.\\
We thank Scott Dhuey at the Molecular Foundry for assistance with the electron beam lithography.\\
We acknowledge the project FISR - C.N.R. “Tecnopolo di nanotecnologia e fotonica per la medicina di precisione” - CUP B83B17000010001 and "Progetto Tecnopolo per la Medicina di precisione, Deliberazione della Giunta Regionale n. 2117 del 21/11/2018.\\
This research is funded in part by the Gordon and Betty Moore Foundation’s EPiQS Initiative, Grant GBMF9615 to L. N. Pfeiffer, and by the National Science Foundation MRSEC grant DMR 1420541.

\bibliographystyle{unsrt}
\bibliography{references}

\newpage

\setcounter{equation}{0}
\setcounter{figure}{0}
\setcounter{table}{0}
\setcounter{page}{1}
\makeatletter
\renewcommand{\theequation}{S\arabic{equation}}
\renewcommand{\thefigure}{S\arabic{figure}}
\pagenumbering{roman}

\noindent
\textbf{\Large Supporting Information}\\

\textbf{Waveguide sample}
The full structure is grown on top of an n$^+$-doped GaAs substrate 500 $\mu$m thick. The cladding layer is made of $500$ nm of Al$_{0.8}$Ga$_{0.2}$As, while the WG is composed by $12$ bilayers of Al$_{0.4}$Ga$_{0.2}$As ($20$ nm thick) and GaAs ($20$ nm thick), and a final bilayer made of $20$ nm of Al$_{0.4}$Ga$_{0.2}$As and $10$ nm of GaAs.\\

\textbf{Gratings fabrication and deposition of ITO}
The pattern transfer of waveguides in GaAs/AlGaAs heterostructures was performed by means of electron beam lithography and subsequent plasma etching. A positive-tone resist ZEP520a 50\% was written using Vistec VB300 on a negative pattern of the designed WG, which allows the formation of the ridges that define each structure and the corresponding gratings. The sample was then developed in amyl acetate and then etched using Oxford PlasmaLab 150 Inductively Coupled Plasma (ICP) Etcher with a chlorine-based recipe. The remaining resist was finally stripped away in dichloromethane.
To enable the application of a filter in the reciprocal space for the residual pumping light, the input and output gratings have different periodicities. The electric contact is obtained by uniformly sputtering 50 nm of Indium Tin Oxide (ITO) on top of the sample; the gratings result completely covered by this layer.\\

\begin{figure*}
    \centering
    \includegraphics[width=1.5\columnwidth]{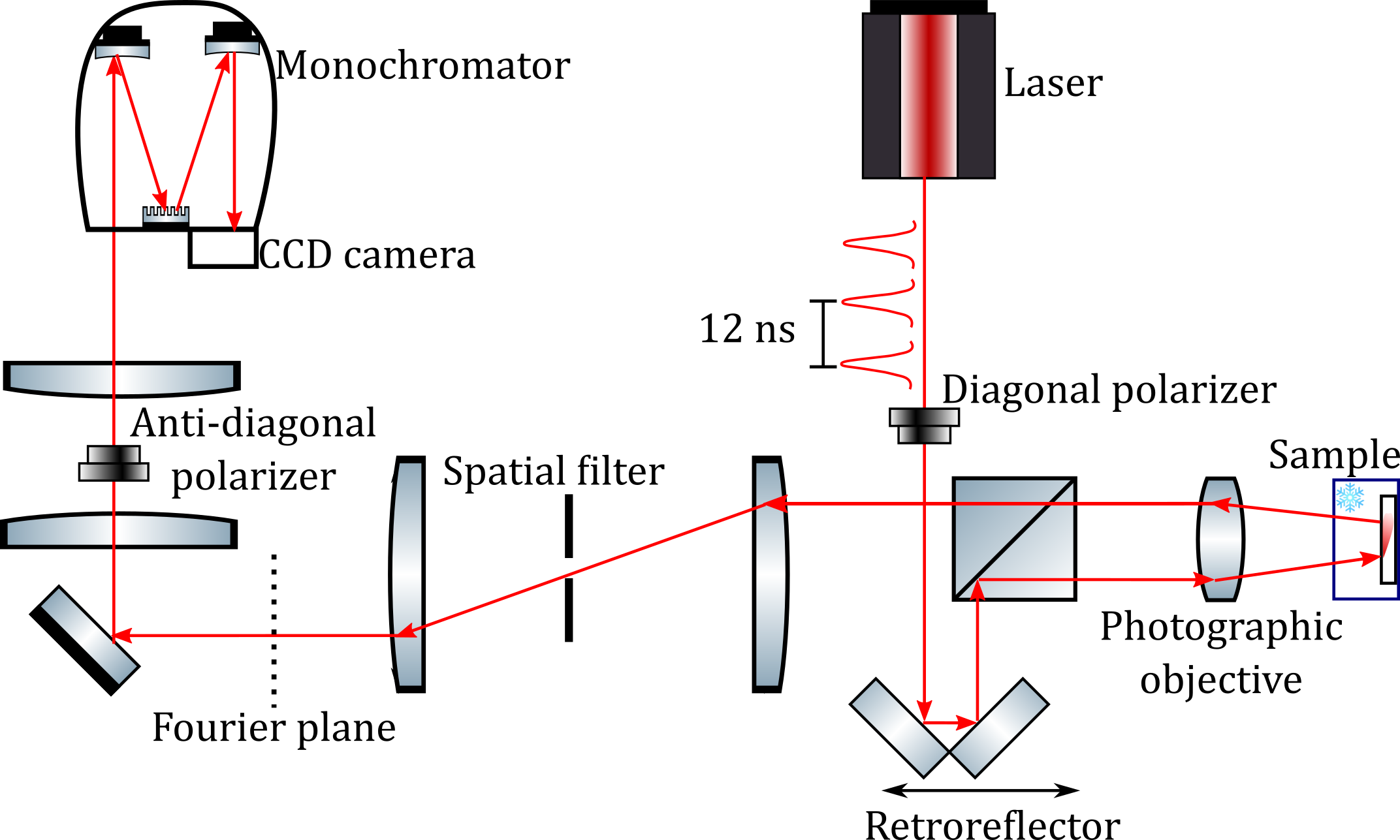}
    \caption{Experimental setup. A ps pulse injects population into the polaritonic waveguide. The transmitted signal is collected and analyzed in space, momentum and energy. A polarization filter and a spatial one are implemented to remove the residual laser signal.}
    \label{sup}
\end{figure*}

\textbf{Optical measurements}
For all the optical characterization the sample is kept at cryogenic temperature of $6$ K. All the PL and resonant measurements are realized in a confocal configuration, using a $5$ ps pulsed laser with a repetition rate of $80$ MHz.
The detection system allows to reconstruct either real or Fourier spaces in a Charge Coupled Device (CCD) coupled to a monochromator $70$ cm long with a diffractive grating with either $600$ or $1800$ lines per mm. This way it is possible to perform measurements resolved in space, angle and energy. An image of the real space is reconstructed before the CCD in order to apply a spatial filter by using a slit, which allows to suppress the residual laser light in both configurations: PL and resonant excitation. For the resonant pump the pulsed laser is tuned at the polariton energy ($\sim1.52$ eV) and it reaches the sample with the right angle to match the linear momentum of the WG mode. To suppress the residual laser in this case, additional to the filter in real space, a polarization filter is implemented in detection. A detailed scheme of the experimental setup is displayed in Fig.~\ref{sup}.

\begin{figure*}
    \centering
    \includegraphics[width=1.0\columnwidth]{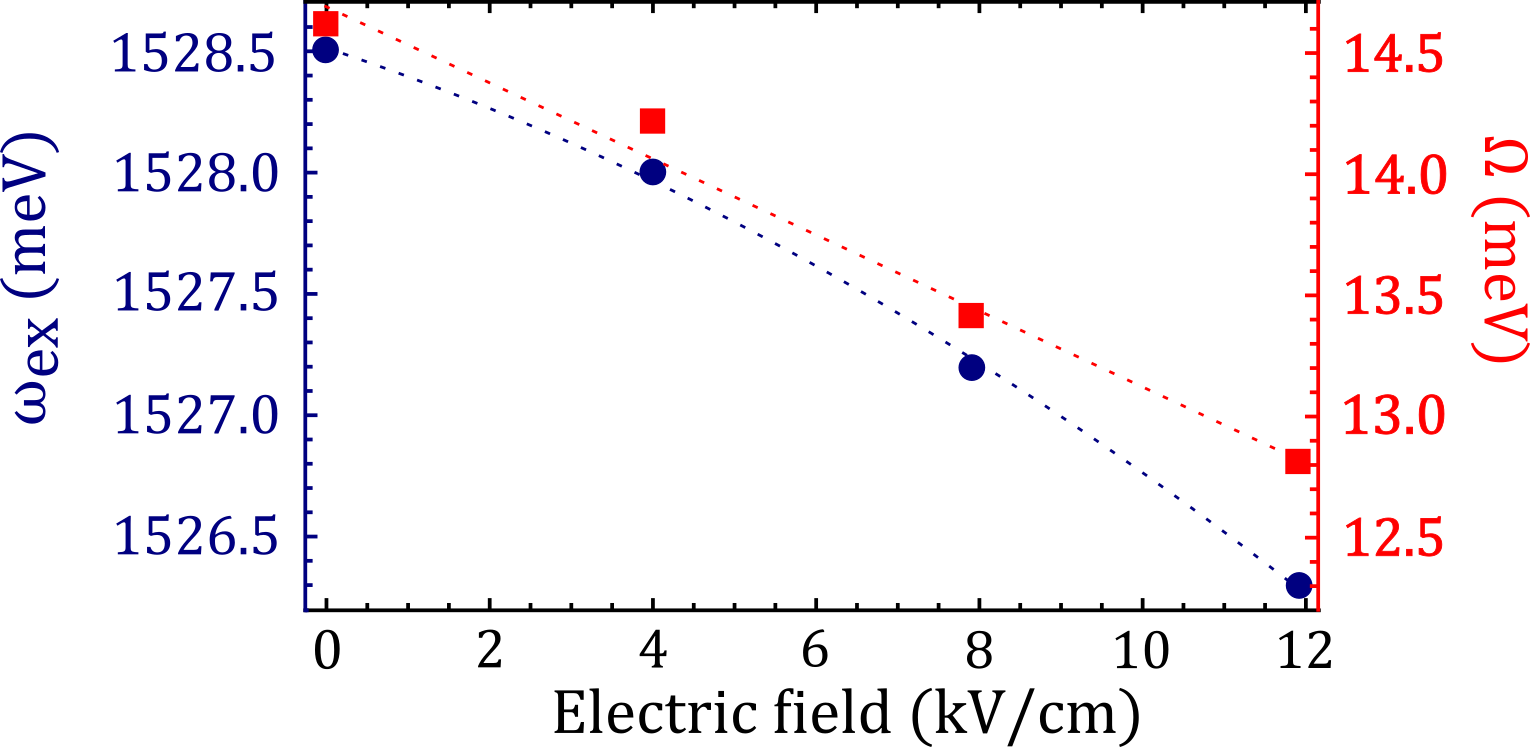}
    \caption{d) Exciton energy (blue circles) and Rabi splitting for the TE mode (red squares) as a function of the applied electric field.}
    \label{stark}
\end{figure*}

\textbf{Stark effect}
Fig. \ref{stark} shows the excitonic energy (blue circles) as a function of the applied electric field. The red squares of Fig.~\ref{stark} show the small reduction in the Rabi splitting due to the spatial separation between electrons and holes.

\begin{figure*}
    \centering
    \includegraphics[width=1.5\columnwidth]{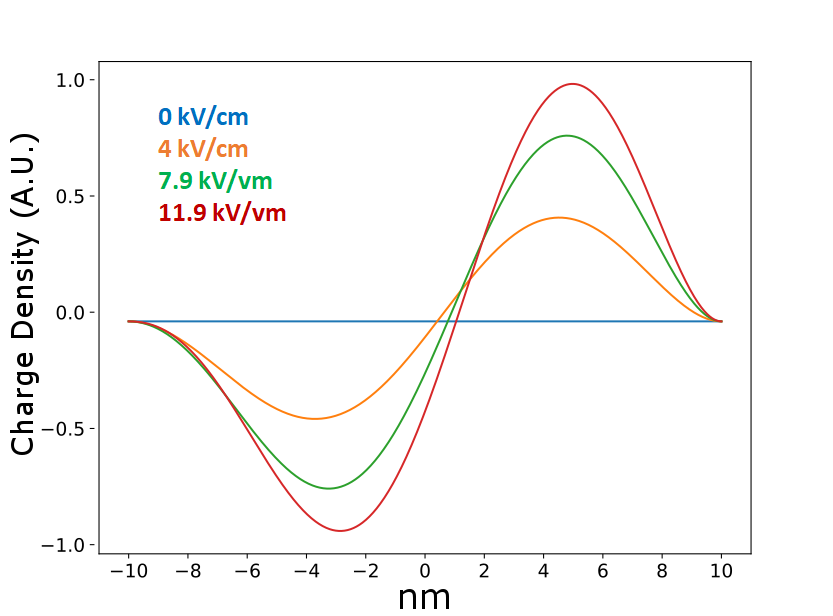}
    \caption{Spatial charge density across the QW, extending from -10 nm to 10 nm, for increasing applied electric fields. From the associated wavefunctions for the hole and the electron it is possible to compute the field-induced dipole length.}
    \label{CD}
\end{figure*}

\textbf{Charge density and dipole length}
The field-induced charge density can be calculated by using a variational approach for describing the electron and the hole wavefunctions under the action of the applied electric field. We use here the infinite well approximation which, for a 20-nm thick well, predicts a shift of the electron and hole energy levels in agreement with the observed Stark shift \cite{Bastard1988, Bastard1983}.
Fig \ref{CD} shows the charge density across a 20-nm QW for for increasing applied electric fields. At zero fields, electrons and holes occupy roughly the same spatial regions, and no net charge density is visible (flat blue line). When the field increases, electrons and holes are spatially separated and a net spatial charge density is formed. 
An estimate of the induced dipole length can be obtained from the wavefunctions, by simply computing the electron and hole mean spatial position. This simple approach gives values of the dipole length comprised between 0 nm for E=0 kv/cm and 4 nm for E = 11.9 kV/cm.  

\textbf{Simulations}
The simulations are done on a two-dimensional, 10 micron by 352 micron grid, with the fourth-order Runge-Kutta algorithm used to advance time. The spatial derivatives are performed using spatial-domain operators with sixth-order accuracy, and the waveguide is implemented by allowing the refractive index and loss constants to be defined independently at each grid point. The input parameters of the simulation are determined experimentally. $\gamma_p$ is 30 $\mu eV$. Parameters that vary with electric field are summarized in the table below.

\begin{center}
\begin{tabular}{ |c|c|c|c| } 
 \hline
 $E ($kV/cm$) $ & Exciton energy ($eV$) & $\Omega$ ($meV$) & $\gamma_e$ ($\mu eV$) \\ 
 \hline
 0 & 1.5285 & 13.9 & 223.24 \\ 
 4.0 & 1.5281&  13.2 & 251.41\\ 
 7.9 & 1.5273& 12.5 & 314.79 \\ 
 11.9 & 1.5265& 12.0 & 350.0 \\ 
 \hline
\end{tabular}
\end{center}

These values of $\gamma_e$ are determined based on the frequency-dependent transmission of the waveguide at low intensity, and are made to be consistent with the input powers and output intensity across all measured frequencies and voltages.

The value of the interaction constant $g$, pump pulse pulse duration, and pump pulse central frequency are determined using the nonlinear least squares algorithm of the GNU Scientific Library. The difference between the power-dependent output spectrum of the simulation and measurement is treated as the error to be minimized. This matrix contains the results 36 independent simulations (at different input powers), and must be solved four times at each iteration of the fitting algorithm. One set of spectra is calculated in approximately 11 seconds on a desktop computer by implementing the solver in both c++ to run on the CPU (8 simulations on 8 cores) and CUDA to run on the GPU (28 simulations). Convergence is typically reached in approximately 10 minutes.\\

\textbf{Calibration of exciton density}
The values of exciton density are estimated based on the simulation results after the aforementioned matching to the experimental data. The value of $\Psi(z,x)$ at the end of the waveguide $(z=z_{m})$ is recorded over the course of the simulation and stored in a 2D matrix $\Psi(x,t)$. An example is shown in the Supplementary Fig. \ref{supPsi}. The maximum value contained in the matrix $|\Psi(x,t)|^2$ is then recorded as the exciton density at the end of the waveguide. 

\begin{figure*}
    \centering
    \includegraphics[width=1.5\columnwidth]{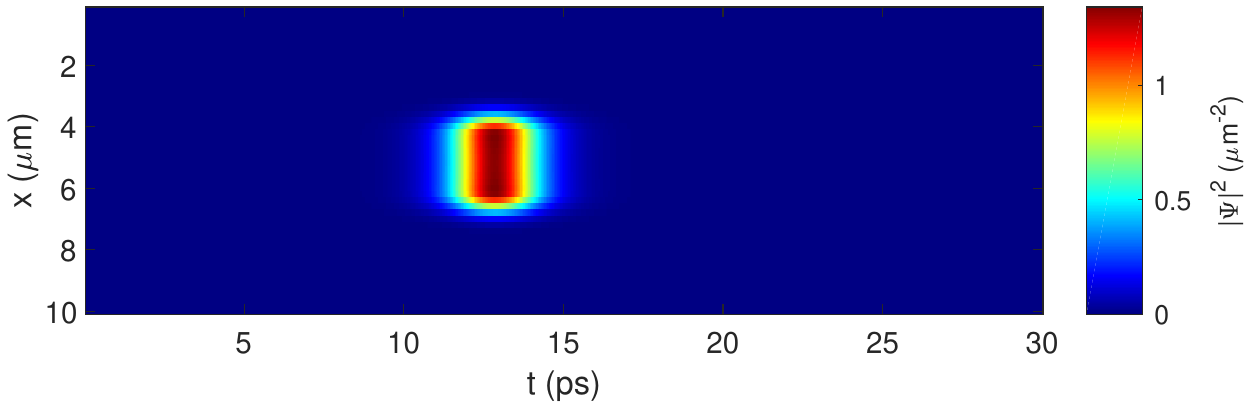}
    \caption{Recorded exciton probability density at the end of the waveguide recorded during the lowest power simulation of the unbiased sample with excitonic fraction 0.5. The estimated exciton density is the peak value of this spatiotemporal matrix.}
    \label{supPsi}
\end{figure*}

We note that in the experiment, the absolute value of $g$ is not obtained, and therefore the value at zero electric field used in the simulation is only an estimation \cite{Estrecho2019}. This carries over to the resulting estimates of the exciton density; deviations of $g$ from the estimated base value of 6 $\mu eV \mu m^2$ will proportionally affect the values of exciton density.

\end{document}